\documentclass[12pt, a4paper]{article}

    \usepackage{cmap} 
\usepackage[T1, T2A]{fontenc}
\usepackage[utf8]{inputenc}
\usepackage[russian,english]{babel}

\usepackage[margin=2cm]{geometry}

\usepackage{microtype}

\usepackage{amsmath, amssymb, amsthm}
\usepackage{indentfirst}
\usepackage[sort,compress]{cite}
\usepackage[affil-it]{authblk}

\usepackage[use-xspace=true]{siunitx}
\usepackage[margin=0.2em]{subcaption}

\usepackage{graphicx}

\usepackage{hyperref}

\DeclareMathOperator*{\argmin}{argmin}
\DeclareMathOperator{\tr}{tr}
\DeclareMathOperator{\bydef}{\overset{\text{def}}{=}}
\DeclareMathOperator{\Expect}{\mathbb{E}}

\newlength{\subfigsize}
\newcommand{\vq}{\mathbf{q}}
\newcommand{\vrho}{\boldsymbol{\rho}}
\newcommand{\vr}{\mathbf{r}}

\newcommand{\prior}{\textnormal{pr}}

\title{Improvement of optical image by measurement reduction technique at parametric multiplexing}
\author[1]{D.\,A.\,Balakin}
\author[1,2]{A.\,S.\,Chirkin}
\affil[1]{M.\,V.\,Lomonosov Moscow State University, Faculty of Physics, Leninskie Gory, 1, bld 2, Moscow 119991, Russia}
\affil[2]{M.\,V.\,Lomonosov Moscow State University, The International Laser Center, Leninskie Gory, 1, bld 62, Moscow 119991, Russia}
\begin{document}

\maketitle

\begin{abstract}
In the process of parametric optical image amplification,
images are formed at new frequencies in addition to the amplified original image.
We show that the parametric multiplexing of optical images
can be used to produce an image with improved quality.
As an example, we study the parametric amplification of an optical image
at low-frequency pumping
in which multiplexed optical images turn out to be quantum-correlated.
Additional improvement is made possible
by using the information about the object that is available to the researcher,
in particular, about sparsity of its image.
To take the available information into account,
we apply the measurement reduction technique.

\end{abstract}

\section*{Introduction}

As it is well-known, in traditional parametric amplification of an optical image
with high-fre\-quen\-cy pumping,
an additional image appears at the so-called idle frequency
(see, for example \cite{Kolobov, Kolobov1}).
In the case of processes of optical parametric amplification
with low-frequency pumping that can be realized in coupled parametric processes  \cite{Morozov,Novikov, chirkin_shutov_2009}
optical images are formed at more than two frequencies \cite{ Chirkin1,Chirkin2, Makeev, Bondani, Allevi }.
In other words, in coupled parametric interactions
frequency multiplexing of optical images occurs.
The quantum theory of such interactions is developed in \cite{ Chirkin1,Chirkin2, Makeev,  Allevi },
where various parametric image amplification schemes are examined.
Experimental studies on the optical image multiplexing
are presented in the articles  \cite{ Bondani, Allevi }.

In the works \cite{ Chirkin1,Chirkin2, Makeev }
optical image quality is characterized by the signal-to-noise ratio (SNR).
It is established that
SNR of the image at the main frequency decreases in spite of increasing the mean photon number.
Meanwhile, SNR of the images at additional frequencies increases
along with their mean photon number.

Recently, image multiplexing has been used in ghost image acquisition schemes
\cite{mgi_icono_lat_2016, mgi_correlations_jrlr_2017, ghost_images_jetp, gi_sparsity_eng, duan_et_al_2013,zhang_et_al_2015,chan_et_al_2009}

It is shown in our works \cite{mgi_icono_lat_2016, mgi_correlations_jrlr_2017, ghost_images_jetp, gi_sparsity_eng}
that using quantum correlations of ghost images
and the reduction technique in image processing,
we can improve the characteristics of the reconstructed optical image.

The purpose of this paper is
to study the quantum correlations of optical images during their multiplexing
in two coupled parametric processes
and apply the method of measurement reduction
for obtaining an optical image with improved quality.
The analysis is based on the coupled parametric interactions
that realize parametric image amplification with low-frequency pumping,
and the scheme with far-away object is considered.
Note that some issues of such a process are studied in \cite{Chirkin2}.

This process is of particular interest
when the wavelength of the original image is in the ultraviolet range
and the pump radiation wavelength of the conventional three-frequency interaction
falls within the absorption region of a nonlinear crystal.
The process of parametric amplification with low-frequency pumping
allows the use of visible radiation as pumping,
while the images at additional frequencies will be in the near-infrared range.

The article structure is as follows.
In section~\ref{sec:amplification}, we discuss the optical setup for image amplification and multiplexing and the parametric processes taking place therein.
In section~\ref{sec:formulation} a specific variant of the optical setup is considered
and the photon number means, variances and covariances are derived.
In section~\ref{sec:image-processing}
the measurement model and the measurement reduction method is outlined,
including the notions of a measuring transducer, an ideal measuring transducer
and the conditions for the possibility of image reconstruction.
The information about the object that is available to the researcher
and that is employed in reduction is summarized in
subsection~\ref{sec:image-information}.
In subsection~\ref{sec:reduction-algorithm},
the algorithm of image processing using reduction method
that takes this information into account
is described.
Computer modeling results are given
in section~\ref{sec:computer-modelling}.
Main results of the article are summarized in the conclusion.

\section{Amplification and multiplexing of an optical image}
\label{sec:amplification}

\begin{figure}
\includegraphics[width=\linewidth]{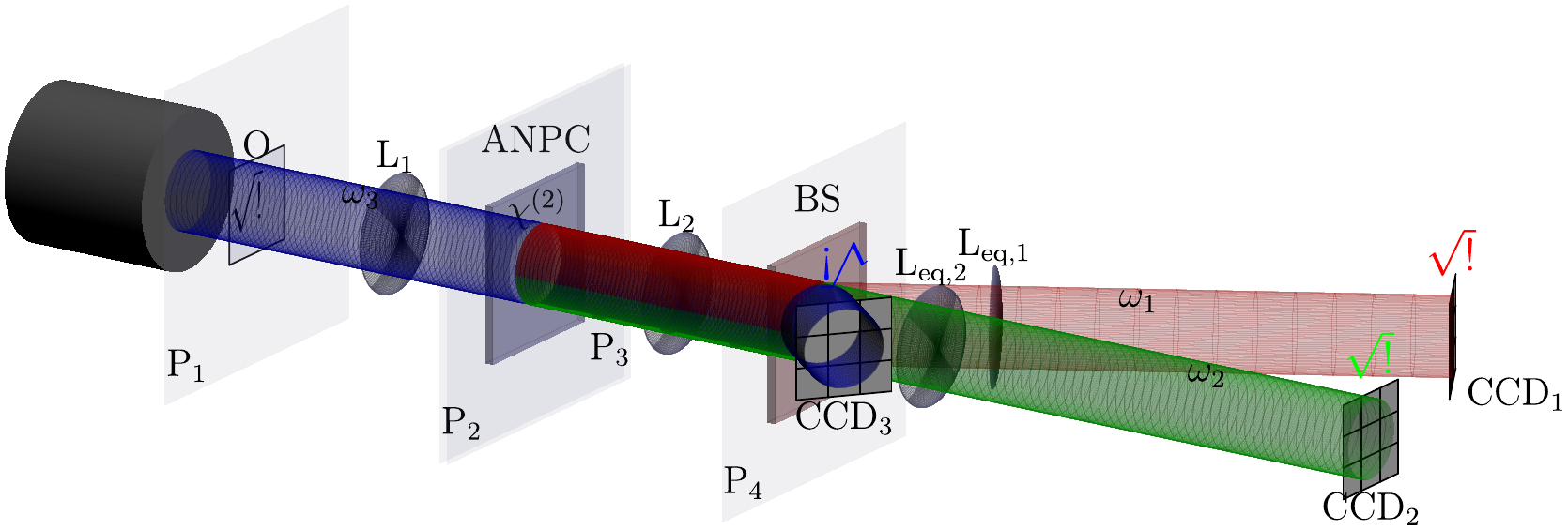}
\caption{The scheme for parametric optical image amplification with far-away object.
The pump at frequency $\omega_3$ illuminates the object O
and the radiation is focused using the lens L$_1$
onto the aperiodic nonlinear photon crystal (ANPC)
where the coupled parametric processes take place
and images at new frequencies are formed.
After ANPC, the radiation is focused by the lens L$_2$ onto the beam splitter (BS),
and is split into beams with frequencies $\omega_1$, $\omega_2$ and $\omega_3$.
The photons with frequencies $\omega_1$ and $\omega_2$ pass through lenses
L$_{\text{eq},1}$ and L$_{\text{eq},2}$, respectively,
to equalize image scales (see sec.~\ref{sec:formulation}).
CCD$_j$ is a CCD camera}
\label{fig:scheme}
\end{figure}

The scheme of parametric optical image amplification is
depicted in Fig.~\ref{fig:scheme}.
The weak optical image that is to be amplified
is located in the object plane $P_1$.
This image is projected by the lens $L_1$ onto the input $P_2$
of an aperidic nonlinear photon crystal (ANPC),
for example, LiNbO$_3$,
in which coupled parametric interactions,
specifically,
down- and up-conversion processes,
occur simultaneously.
The amplified image and the images generated at two new frequencies (see below)
are projected by the lens $L_2$ from the output of the crystal (plane $P_3$)
onto the image plane $P_4$.
The lenses $L_1$ and $L_2$ have the same focal length $f$.
The object and image planes, as well as the input and the output of the ANPC,
are at the distance $f$.
This is the scheme with the so-called far-away object,
it is similar to the scheme considered in \cite{Kolobov1, Chirkin2}.

We denote the field operators in the object and the image planes
as $A_{j0}(\vrho)$ and $A_j(\vr)$, respectively,
and those in the input and the output planes of the NPC
as $A_j^\textnormal{in}(\vr_1)$ and $A_j^\textnormal{out}(\vr_2)$, respectively.
The index $j$ is associated with the wavelength $\lambda_j$ .
The operators $A_{j0}$, $A_j^\textnormal{in}$ and $A_j^\textnormal{out}$, $A_j$ are
related by the Fourier transformation
performed by the lens $L_1$
\begin{equation}\label{eq: b_in}
A_{j}^\textnormal{in}(\vr_1) =
\frac{1}{\lambda_{j} f} \int\limits_{-\infty}^\infty A_{j0}(\vrho)
\exp\left(-i\frac{2\pi}{\lambda_j f}\vrho\vr_1\right) d\vrho,
\end{equation}
and by the lens $L_2$
\begin{equation}\label{eq: e_in}
A_{j}(\vr) =
\frac{1}{\lambda_{j} f} \int\limits_{-\infty}^\infty
\left[  A_{j}^\textnormal{out}(\vr_2) P(\vr_2)
+ (1-P^2(\vr_2))^{1/2} \hat{v} (\vr_2)\right]
\exp\left(-i\frac{2\pi}{\lambda_j f}\vr\vr_2\right)
d\vr_2.
\end{equation}
$P(\vr_2)$ is the pupil frame function that accounts
for the finite area $S_a$ of the pupil.
Taking it into account is necessary
for the correct analysis of the image amplification scheme \cite{Kolobov1}
due to vacuum fluctuations outside of the pupil's aperture,
described by the operator $ \hat{v} (\vr_2)$ (the second term in Eq.~\eqref{eq: e_in}).
However, they do not contribute
to any normal-ordered operational expressions associated with measurable values,
so we omit this term below.
Therefore, the expression \eqref{eq: e_in} can be presented as
\begin{equation}\label{eq: e_in2}
A_{j}(\vr) =
\frac{(2\pi)^{2}}{\lambda_{j} f}
\int\limits_{-\infty}^\infty
a_{j}^\textnormal{out}(\vq) P\left(\vq-\frac{k_{j}}{f}\vr\right)
d\vq,
\end{equation}
where $a_{j}^\textnormal{out}(\vq)$ and $P(\vq)$ are the Fourier transforms of $ A_{j}^\textnormal{out}(\vr) $ and $P(\vr)$, respectively:
\begin{equation}\label{eq: fourier}
a_j(\vq) =
\frac{1}{(2\pi)^2}
\int\limits_{-\infty}^\infty
A_j(\vr) e^{-i\vq\vr}d\vr,
\end{equation}
 $\vq$ is the transversal wave vector.

All operators under consideration
$A_{j0}(\vr)$, $A_j(\vr)$, $A_j^\textnormal{in}(\vr)$ and $A_j^\textnormal{out}(\vr)$
obey the commutation relations
\begin{equation}\label{eq: commutation}
[A(\vr, z), A^\dag(\vr', z)] = \delta(\vr - \vr'),
\quad
[A(\vr, z), A(\vr', z)] = 0,
\end{equation}
where $z$ is the direction of wave propagation.
Mean value $\langle\hat{N}(\vr,z)\rangle$
of the operator $\hat{N}(\vr, z) = A^\dag(\vr, z) A(\vr, z)$
is the mean photon flux density in cross-section $z$, measured in photons per cm$^2$.

To find the connection between operators
$A_j^\textnormal{in}(\vr_1)$ and $A_j^\textnormal{out}(\vr_2)$,
i.\,e. the connection between fields at the output and the input of the ANPC,
it is necessary to consider the nonlinear processes in the ANPC.
The processes under study are
two coupled processes
\begin{equation}\label{eq: processes}
  \begin{array}{l}
    \omega_p=\omega_1+\omega_2, \\
    \omega_p+\omega_1=\omega_3.
  \end{array}
\end{equation}
Here $\omega_p$ is the frequency of intense pump wave,
and $\omega_1$, $\omega_2$ and $\omega_3$
are the frequencies of generated waves
with $\omega_p$ and $\omega_1$ being the shared frequencies of two processes.
The first down-conversion process in Eq.~\eqref{eq: processes}
represents parametric amplification during high-frequency pumping,
and the second one is the up-conversion process.
They can be implemented simultaneously in an aperiodical NPC. 

In the undepleted pump plane wave approximation
taking into account the diffraction phenomenon
the processes in Eq.~\eqref{eq: processes} can
be described by the system of equations
\begin{equation}
\label{eq:diffraction_system}
  \left\{
    \begin{array}{l}
      \displaystyle\frac{\partial A_1}{\partial z}-\frac{i}{2k_1}\triangle_\perp A_1 =
      i\beta A_2^\dag+i\gamma A_3, \\
      \displaystyle\frac{\partial A_2}{\partial z}-\frac{i}{2k_2}\triangle_\perp A_2 =
      i\beta A_1^\dag, \\
      \displaystyle\frac{\partial A_3}{\partial z}-\frac{i}{2k_3}\triangle_\perp A_3 =
      i\gamma A_1.
    \end{array}
  \right.
\end{equation}
Here $\triangle_{\perp} = \triangle_{\perp}(x,y)$ is the transversal Laplacian,
$A_j^{\dag} = A_j^{\dag}(\vr, z)$ and $A_j = A_j(\vr, z)$
are the creation and annihilation operators of photons
with frequency $\omega_j$ ($j = 1, 2, 3$) respectively,
$\beta$ and $\gamma$ are real nonlinear coupling coefficients
that are proportional to second order nonlinear susceptibility
and the absolute value of pump wave amplitude \cite{Morozov}.
Eqs.~\eqref{eq:diffraction_system} are derived for a lossless ANPC
and for interaction of monochromatic waves.

The system of Eqs.~\eqref{eq:diffraction_system} is solved
by applying Fourier transform
\begin{equation}\label{eq: Fourier}
A_j(\vr, z) =
\int\limits_{-\infty}^\infty a_j(\vq, z)e^{i \vq\vr}d\vq,
\end{equation}
after which the Eqs.~\eqref{eq:diffraction_system} become
\begin{equation}\label{eq: Fourier
system}
  \left\{
    \begin{array}{l}
      \displaystyle\frac{da_1}{dz}=-i\mu_1 a_1+i\beta a_2^\dag+i\gamma a_3, \\
      \displaystyle\frac{da_2^{\dag}}{dz}=i\mu_2a_2{^\dag}-i\beta a_1, \\
      \displaystyle\frac{da_3}{dz}=-i\mu_3 a_3 +i\gamma a_1,
    \end{array}
  \right.
\end{equation}
where $a_j = a_j(\vq, z)$, $\mu_j=\frac{q^2}{2k_j}$.

In the matrix form the solution of Eqs. (\ref{eq: Fourier system})
has the form
\begin{equation}\label{eq: matrix solution}
\mathbf{a} = Q \mathbf{a}_0,
\end{equation}
where $\mathbf{a}_0^T=(a_{10},a_{20}^\dag,a_{30})$ is determined by
the values of the operators at ANPC input ($z = 0$),
the index $T$ denotes transposition.
In the case considered below,
the operators $\mathbf{a}_{10}(\vq)$, $\mathbf{a}_{20}(\vq)$ describe the vacuum state
and the operator $\mathbf{a}_{30}(\vq)$ describes a coherent state.

The matrix $Q$ consists of transfer functions $Q_{nm}$:
\begin{equation}
\label{eq:Q}
Q =
\begin{pmatrix}
Q_{11} & Q_{12} & Q_{13} \\
Q_{21} & Q_{22} & Q_{23} \\
Q_{31} & Q_{32} & Q_{33}
\end{pmatrix}.
\end{equation}
The functions $Q_{nn} = Q_{nn}(q, z)$ are the self-transfer functions
because they describe the amplification at frequencies $\omega_n$ ($n=1, 2, 3$),
while the cross-transfer functions $Q_{nm} = Q_{nm}(q, z)$ describe
the conversion from frequency $\omega_m$ to $\omega_n$.
Elements of the matrix \eqref{eq:Q} can be found in \cite{Chirkin2}
and are not given here due to being cumbersome.

\section{Formulation of the problem}
\label{sec:formulation}
 
In the previous section,
the quantum theory of two coupled parametric processes is presented
in relation to amplification and frequency conversion of an optical image,
which can arrive at the nonlinear crystal
at any of the frequencies $\omega_1$, $\omega_2$ or $\omega_3$.
Here we turn to the case when
the image with a mean photon number density
$\left\langle \hat{N}_{30}(\vr) \right\rangle$
is fed to the crystal in a coherent state at the frequency $\omega_3$.
In the framework of the monochromatic waves under consideration,
results given below are valid
if the image registration time is less than the characteristic time of image change,
for example, the correlation time.
 
In the image plane $P_4$, the mean photon number over pixel area $S_p$
in the amplified image and the additional ones is given by the expressions 
\begin{equation}\label{eq:mean_photons}
  \begin{array}{l}
    \left\langle  \hat{N}_{3}(\vr)\right\rangle = S_p |Q_{33}(k_{3}r/f )|^2  \left\langle  \hat{N}_{30}(-\vr)\right\rangle , \\
    \left\langle  \hat{N}_{2}(\vr)\right\rangle = (\lambda_3/\lambda_2)^{2} S_p |Q_{23}(k_{2}r/f)|^2  \left\langle  \hat{N}_{30}(-(\lambda_3 /\lambda_2) \vr)\right\rangle,  \\
    \left\langle  \hat{N}_{1}(\vr)\right\rangle = (\lambda_3/\lambda_1)^{2} S_p |Q_{13}(k_{1}r/f)|^2  \left\langle  \hat{N}_{30}(-(\lambda_3 /\lambda_1) \vr)\right\rangle
  \end{array}
  \end{equation}
As expected, the output optical images are inverted relative to the initial image.
It is important to note that
the scales of output images at different frequencies are different,
and the change in spatial scale is determined
by the coefficient $\lambda_{3} / \lambda_{j} $,
where $\lambda_{j} $ is the image wavelength.
It should be noted that this fact was not taken into account in \cite{Chirkin2}.

Different image scales somewhat complicate the reduction algorithm,
as this means that sizes of pixels are different in different arms.
One can proceed further in several ways.
\begin{itemize}
\item In the general approach without the assumption
that the image is piecewise constant (see sec.~\ref{sec:image-processing}),
dealing with different pixel sizes in different arms is avoided,
since in the infinite-dimensional case pixel size does not affect image representation.
\item The images can be rescaled during processing
if sensor point spread functions allow to do this
both accurately and without loss of data.
\item Finally, one can bring the images to the same scale using additional lenses.
This approach is considered below.
\end{itemize}
It should be noted that these approaches provided the same results when they are valid.

In the image plane $P_4$,
images with different frequencies are directionally separated
(for example, using a prism).
A lens with the focal distance $f_j$ ($j = 1, 2$)
is placed on the path of radiation with frequency $\omega_j$
in order to bring the image to the spatial scale
that coincides with the scale of the image at the frequency $\omega_3$.
The lens is located at a distance $l_{j1}$ from the plane $P_4$
and a distance $l_{j2}$ to the measurement plane.
The specified distance must satisfy the lens formula
\begin{equation}\label{eq: lens}
\frac{1}{l_{j1}} + \frac{1}{l_{j2}}=\frac{1}{f_j}.
\end{equation}
In this case, the distribution of the image mean photon number
in the optically conjugate plane is given by ($j=1, 2$):
\begin{equation}
\label{eq:trans}
\left\langle\hat{N}_{j}^{tr}(\vr)\right\rangle =
\left(\frac{l_{j1}}{l_{j2}}\right)^{2} \left\langle  \hat{N}_{j}\left(-\frac{l_{j1}}{l_{j2}}\vr\right)\right\rangle =
\left(\frac{\lambda_3 l_{j1}}{\lambda_{j} l_{j2}}\right)^{2} S_p
\left| Q_{j3} \left( \frac{k_{2}r l_{j1}}{f l_{j2}}\right)\right|^2
\left\langle
\hat{N}_{30}\left(  \frac{\lambda_3 l_{j1}}{ \lambda_jl_{j2}} \vr\right)\right\rangle.
\end{equation}
According to Eq.~\eqref{eq:trans}, to match the image scales at the wavelengths
$\lambda_j$ and $\lambda_3$ the ratio 
\begin{equation}
\label{eq: lens2}
\frac{l_{j1}}{l_{j2}} = \frac{\lambda_j}{\lambda_3}.
\end{equation} 
must be satisfied. Under this condition, we have 
\begin{equation}
\label{eq: trans2}
\left\langle \hat{N}_{j}^{(tr)}(\vr) \right\rangle =
S_p \left|Q_{j3}\left(\frac{k_{3}r }{f }\right)\right|^2
\left\langle  \hat{N}_{30}(\vr)\right\rangle. 
\end{equation}
The photon number variances
$\sigma^2_j =
\left\langle \hat{ N^{2}_{j}}(\vr)\right\rangle
- \left\langle \hat{N}_{j}(\vr)\right\rangle^2$
are given by the following formulas
\begin{equation}\label{eqn:photocount-variance}
\begin{array}{l}
\sigma^2_3 = S_p \left[ 1+2  |Q_{32}(k_{3}r)/f)|^2 \right] |Q_{33}(k_{3}r/f)|^2 \left\langle  \hat{N}_{30}(-\vr)\right\rangle, \\
\sigma^2_2 =  S_p \left[ 1+2 \frac {S_{a}S_p}{(f\lambda_3)^{2}} |Q_{21}(k_{3}r/f)|^2 \right] |Q_{23}(k_{3}r/f)|^2 \left\langle  \hat{N}_{30}(\vr)\right\rangle,  \\
\sigma^2_1 =  S_p \left[ 1+2  |Q_{12}(k_{3}r/f)|^2 \right] |Q_{13}(k_{3}r/f)|^2 \left\langle  \hat{N}_{30}( \vr)\right\rangle.
\end{array}
\end{equation}

Finally, the mutual correlations of the image fluctuations (covariances) between   different frequencies 
\begin{equation}\label{eq: covar}
  C_{jl}(\vr)= \left\langle \hat{N}_j(\vr)\hat{N}_l(\vr)\right\rangle-\left\langle \hat{N}_j(\vr)\right\rangle \left\langle \hat{N}_l(\vr)\right\rangle
\end{equation}
have the forms
\begin{equation}\label{eq: covar_1}
C_{31}(  \vr)=C_{13}(  \vr) = \frac{S_a S_p^2}{(f\lambda_3)^2} \left\lbrace   |Q_{12}(k_{3}r/f)  Q_{33}(k_{3}r/f)|^2
+  |Q_{13}(k_{3}r/f) Q_{32}(k_{3}r/f)|^2 \right\rbrace   \left\langle  \hat{N}_{30}( \vr)\right\rangle,
\end{equation}
\begin{equation}\label{eq: covar_2}
C_{32}(  \vr)=C_{23}( \vr) = \frac{S_a S_p^2}{(f\lambda_3)^2}  |Q_{23}(k_{3}r/f)|^2
\left\lbrace | Q_{32}(k_{3}r/f)|^{2}
+  | Q_{33}(k_{3}r/f)|^2 \right\rbrace   \left\langle \hat{N}_{30}(\vr)\right\rangle,
\end{equation}
\begin{equation}\label{eq: covar_3}
C_{12}(  \vr)=C_{21}(  \vr)=\frac{S_a S_p^2}{(f\lambda_3)^2}   |Q_{23}(k_{3}r)/f)|^2 \left\lbrace |Q_{12}(k_{3}r/f)|^2
+  |Q_{13}(k_{3}r)/f)|^2 \right\rbrace  \left\langle  \hat{N}_{30}(\vr)\right\rangle,
\end{equation}

\begin{figure}[ht]
\centering
\includegraphics[scale=1]{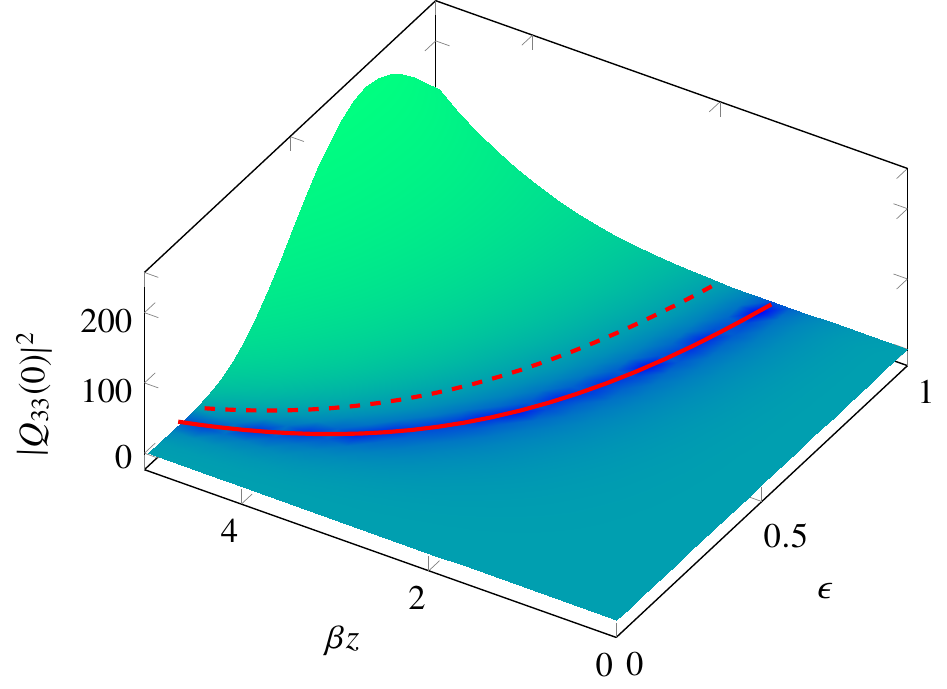}
\caption{$|Q_{33}(0)|^2$ as a function of crystal parameter $\epsilon = \gamma/\beta$ and the dimensionless crystal length $\beta z$.
The solid line shows $\beta z_0$ (disappearance of the amplified image)
as a function of $\epsilon$
and the dashed line shows $\beta z_m$ (unit amplification)
as a function of $\epsilon$}
\label{fig:q33}
\end{figure}

To estimate the above moments, we use their values at $\vr = 0$.
In this case, the matrix elements \eqref{eq:Q} have a simple analytical form:
\begin{equation}\label{eq:matrix_3}
\begin{array}{l}
Q_{12}(0) = -Q_{21}(0) = i(\beta/\Gamma) \sinh \Gamma z ,\\
Q_{13}(0) = Q_{31} (0) = i(\gamma/\Gamma) \sinh \Gamma z ,\\
Q_{23}(0) = -Q_{32}(0) = (\beta \gamma/\Gamma^2) (\cosh \Gamma z - 1), \\ 
Q_{33}(0) = 1 -  (\gamma/\Gamma)^2(\cosh \Gamma z - 1),  
\end{array}   
\end{equation}   
where $\Gamma=(\beta^2 - \gamma^2)^{1/2}$.
$|Q_{33}(0)|^2$ as a function of $\gamma/\beta$ and $\beta z$ is shown in Fig.~\ref{fig:q33}.

It follows from Eq.~\eqref{eq:matrix_3} that
the amplification of image takes place if $\beta > \gamma$.

In the absence of a down-conversion process ($\beta = 0$)
and, therefore, without generation of additional frequencies,
there is no amplification.
In the coupled parametric process under the condition $\beta > \gamma$,
the original image initially decays and at the interaction length $z_0$,
$\cosh z_0 = (\beta/\gamma)^2$,
the mean photon number $\langle \hat{N}_{30}(\vr) \rangle$ of the image becomes zero.
The image gain process begins after the interaction length $z_m$,
$\cosh z_m = 2 (\beta/\gamma)^2 - 1$ (see also Fig.~\ref{fig:q33}).
As for the photon numbers
$\langle \hat{N}_1(\vr)\rangle$, $\langle \hat{N}_2(\vr)\rangle$
at frequencies $\omega_1$ and $\omega_2$,
they monotonously grow as the interaction length increases
according to expressions \eqref{eq:mean_photons}, \eqref{eq:matrix_3}.

\section{Processing of acquired images}
\label{sec:image-processing}

The output of sensors in the $i$-th arm, denoted as $\xi^{(i)}(\vr)$,
can be considered as the output of a measuring transducer (MT)
for input signal $g(\vr) \sim \langle \hat{N}_{30}(-\vr) \rangle$.

We will consider piecewise constant images,
i.\,e. transparency of the research object is constant within each pixel.
Areas of constant transparency and constant brightness corresponding to pixels
are considered to be ordered in an arbitrary but fixed way.
Due to that it is sufficient for us to consider a finite number of values of $\vr$.
Thus, $g$ as the vector of transparencies is an element of
finite-dimensional Euclidean space $\mathcal{F}$.

This assumption is made for simplicity
(in order to avoid working with infinite-dimensional spaces)
and is not crucial to the reduction method.
For examples of reduction in the infinite-dimensional case
see, e.\,g., \cite[ch.~10]{pytyev_ivs} and \cite{chulichkov_vmu}.
If the assumption is invalid, the estimate of the algorithm below
estimates linear combinations of $\langle \hat{N}_{30}(-\vr) \rangle$
around the values of $\vr$ determined by locations of the sensors
with weights that are dependent on the MT and the ideal MT $U$
specified by the researcher (see below).
For example,
if sensors have uniform light sensivity throughout their area,
the ideal MT corresponds to unit-sized sensors of unit size
and the images were acquired by sensors that are several times larger,
$g$ is averaged over unit-sized areas centered at sensor locations.

Let us formulate the measurement model as
\begin{equation}
\label{eqn:measurement-model}
\xi = A g + \nu,
\end{equation}
where $g$ is an unknown vector that describes the transparency distribution
of the object,
$\nu$ is the measurement error with zero expectation, $\Expect \nu = 0$,
which means absence of systematic measurement error,
and covariance matrix
$\Sigma_{\nu} = \Expect \nu \nu^*$.
The dimension of vector $g$ is the number of pixels in the object image,
while the dimension of $\xi$ is the total number of sensors.
The condition of systematic measurement error absense
$\Expect \nu = 0$
means, in particular, that the expectation of the component of measurement results
caused by sensor dark noises is subtracted from the measurement results,
similar to \cite{gi_compressed_sensing_substr_const, gi_sparsity_eng} for ghost images.

The matrix $A$ describes image amplification, multiplexing and acquisition:
the matrix element $A_{ij}$ is equal to the mean output of $i$-th sensor
for unit transparency of $j$-th element of the illuminated object and zero transparency of other object elements (i.\,e. whose indices differ from $j$).
Due to the measuring setup with three arms,
it is a block matrix and consists of three blocks
describing different arms:
\begin{equation}
\label{eqn:a-op-form}
A = \begin{pmatrix}
B_{1} C_{1}\\
B_{2} C_{2}\\
B_{3} C_{3}
\end{pmatrix}.
\end{equation}
Under the conditions used to derive
the mean numbers of photons, their variances and mutual correlations
the matrices $C_1$--$C_3$ are identity ones multiplied by
the factor before $\langle \hat{N}_{30}(-\vr) \rangle$
in expression~\eqref{eq: trans2} for the mean numbers of photons.
The matrices $B_1$--$B_3$ model the sensors.
Specifically, the matrix element $(B_i)_{pk}$ is equal to
the output of the sensor in $i$-th arm at $p$-th position
for unit brightness of $k$-th pixel of the image formed in that arm
and zero brightness of other pixels.
For the same reason, the noise covariance matrix has block form as well:
\begin{equation}
\label{eqn:sigma-op-form}
\Sigma_\nu =
\begin{pmatrix}
B_1 \Sigma_{11}(g) B_1^* &
B_1 \Sigma_{12}(g) B_2^* &
B_1 \Sigma_{13}(g) B_3^*\\
B_2 \Sigma_{21}(g) B_1^* &
B_2 \Sigma_{22}(g) B_2^* &
B_2 \Sigma_{23}(g) B_3^*\\
B_3 \Sigma_{31}(g) B_1^* &
B_3 \Sigma_{32}(g) B_2^* &
B_3 \Sigma_{33}(g) B_3^*
\end{pmatrix}
+ \Sigma_{\nu'}(g).
\end{equation}
Here the element with indices $k$, $k'$
of the block~$\Sigma_{ij}$ is equal to
either photocount variance in Eq.~\eqref{eqn:photocount-variance} if $i = j$
or covariance of photocounts in Eq.~\eqref{eq: covar_1}--\eqref{eq: covar_3} if $i \neq j$
for the same pixel ordering as in the matrix $\mathbf{A}$.
Hence, the dependence of the matrix \eqref{eqn:sigma-op-form} on $g$
is caused by the dependence of variances and covariances on $\langle \hat{N}_{30}(-\vr) \rangle$.
The term $\Sigma_{\nu'}$
is the covariance matrix of the noise component $\nu'$
that is unrelated to parametric amplification and multiplexing,
e.\,g. thermal noise in circuits, detection of outside photons, non-unit quantum efficiency of the sensors and their dark noise.


The objective of an image processing algorithm
is to output the most accurate estimate
of the signal $U g$ from the measurement result $\xi$,
where the matrix $U$ describes
a measuring device that is ideal (for the researcher).
Hence, $U g$ is the feature of the original image $g$
that is of interest to the researcher.
We consider the case when the researcher is interested in
reconstruction of the object image itself,
and imaging does not distort the object,
therefore, $U = I/n$, where $n$ is the average number of photons per pixel of the illuminated object.
One way of achieving this is the measurement reduction method
described in \cite{pytyev_ivs}, see also \cite{pytyev_chulichkov_1998, pytyev_et_al_2004, pytyev_2010, reduction_vmu}.
If the estimation process is described by a linear operator $R$
($R \xi$ is the result of processing the measurement $\xi$),
the corresponding mean squared error (MSE)
in the worst case of $g$,
$h(R, U) = \sup\limits_{g \in \mathcal{F}} \Expect \lVert R \xi - U g \rVert^2$,
as shown in \cite{pytyev_ivs},
is minimal for $R$ that is equal to the linear unbiased reduction operator
\begin{equation}
\label{eqn:reduction-operator}
R_* \bydef U (A^* \Sigma_{\nu}^{-1} A)^- A^* \Sigma_{\nu}^{-1},
\end{equation}
where ${}^-$ denotes pseudoinverse.
$h(R_*, U) = \tr U (A^* \Sigma_{\nu}^{-1} A)^{-1} U^*$,
and the covariance matrix of the linear reduction estimate $R_* \xi$
is
\begin{equation}
\label{eqn:reduction-cov-op}
\Sigma_{R_* \xi} = U (A^* \Sigma_{\nu}^{-1} A)^{-1} U^*.
\end{equation}

Estimation is possible (MSE is finite)
if the condition $U (I - A^- A) = 0$ holds,
where, as noted above, $A$ characterizes the \emph{real} measuring device,
while $U$ characterizes an \emph{ideal} one
with the point spread function required by the researcher,
and, therefore, \emph{the desired resolution},
if this condition if fulfilled.
This condition essentially means that only the features of the object
that are measured by the real measuring device
(that is, affect its output)
can be estimated.
Unlike fluorescence-based superresolution techniques,
see e.\,g. \cite{solomon_et_al_2018},
the proposed technique does not require attaching fluorescent molecules to the object.
However, in addition to the above condition,
the error of the obtained estimate can be too large
to distinguish the signal from the noise in practice,
and usually the better the desired resolution of the ideal measuring device
compared to the resolution of the real one,
the larger MSE of the obtained estimate.
Nevertheless, by choosing $U$ one can select an acceptable (to him) compromise
between obtained resolution and noise magnitude,
that is, to estimate with acceptable resolution
and with an tolerable noise level.

In the case under consideration, as seen from Eq.~\eqref{eqn:a-op-form},
the diagonal elements of $C_1$--$C_3$ are nonzero.
Therefore, each block of $C_j$ is non-degenerate,
so for non-degenerate $B_j$ the reduction error takes only finite values.

The measurement reduction technique for the case when
it is known that $u = U g \in \mathcal{U}_{\prior} \subset \mathcal{U}$,
where $\mathcal{U}_{\prior}$ is convex and closed,
in other words,
when the feature of interest of the object
is known to satisfy certain given constraints,
was considered in \cite{reduction_vmu, lomo_readings}.
The linear estimate \eqref{eqn:reduction-operator}
is refined using this information
by solving the equation
\begin{equation}
\label{eqn:reduction-estimate-in-set}
\hat{u} = \Pi_{\Sigma_{R_* \xi}}\left(\tilde{R}_{\Sigma_{R_* \xi}} \left( \xi^T, \hat{u}^T \right)^T\right)
\end{equation}
for $\hat{u}$,
where
$\tilde{R}_{\Sigma_{R_* \xi}}$ is the measurement reduction operator
for a MT $\left(A^T, U^T\right)^T$
and noise with covariance matrix $\begin{pmatrix}\Sigma_{\nu} & 0\\ 0 & \Sigma_{R_* \xi}\end{pmatrix}$,
and the operator
\begin{equation}
\label{eqn:mahalanobis-projection}
\Pi_{\Sigma_{R_* \xi}}(u) \bydef
\argmin\limits_{v \in \mathcal{U}_{\prior}} (v - u, {\Sigma_{R_* \xi}}^{-1} (v - u))
\end{equation}
describes orthogonal projection onto $\mathcal{U}_{\prior}$
by minimizing the Mahalanobis distance
$\lVert \Sigma_{R_* \xi}^{-1/2} \cdot \rVert$
associated with the covariance matrix
$\Sigma_{R_* \xi}$ \eqref{eqn:reduction-cov-op}
of the linear reduction estimate $R_* \xi$.
The earlier version of reduction technique proposed in \cite{ghost_images_jetp} and in \cite{reduction_vmu}
for similar information
used minimization of the ``ordinary'' Euclidean distance
instead of Mahalanobis distance.
In \cite{lomo_readings},
the advantages of minimizing Mahalanobis distance instead of Euclidean distance
during projection are shown.
For such prior information, the covariance matrix \eqref{eqn:reduction-cov-op}
of linear reduction estimate error
is an upper bound on the covariance matrix of the obtained estimate $\hat{u}$.

\subsection{Prior object information}
\label{sec:image-information}

It is obvious that a priori the transparency distribution of the object
takes values in $[0, 1]$,
hence $g \in [0, n]^{\dim \mathcal{F}}$,
$U g \in [0, 1]^{\dim \mathcal{F}}$,
where the average number $n$ of photons per pixel of the illuminated object
is assumed to be known.

It is assumed that the transparency distribution of the object is not ``entirely'' arbitrary:
transparencies of neighboring pixels usually do not differ much.
As a result, the image is sparse (many of its components are zero)
in a given basis,
similarly to compressed sensing
\cite{candes_wakin_2008, chan_lu_2014, imaging_small_n_photons, mertens_et_al_2017}.
The hypothesis ``$i$-th component in the given basis of the estimate $\hat{u}$ is zero''
is treated as a statistical hypothesis
that is tested using the measurement data
against the alternative that it is nonzero.
Its testing is controlled by choosing the significance level
(the probability of rejecting the hypothesis when it is true)
of the rejection criterion
or a parameter $\tau$ of the criterion that monotonously depends on it.

The researcher also knows the matrix $A$ \eqref{eqn:a-op-form}
that describes image acquisition conditions
and, up to the vector $g$, the matrix $\Sigma_{\nu}$ \eqref{eqn:sigma-op-form}
that describes the magnitudes of measurement errors.
Note that the worst case of $g$ is realized if all pixels are equally transparent.

\subsection{Reduction algorithm}
\label{sec:reduction-algorithm}

The proposed algorithm of
multiplexed GI processing using measurement reduction technique
that is based on the indicated prior information
has the following form.
\begin{enumerate}
\item
\label{itm:first-estimate}
Calculation of linear reduction estimate $R_* \xi$ \eqref{eqn:reduction-operator}
based on the acquired images.
For calculation of the covariance matrix \eqref{eqn:sigma-op-form},
the worst case, that all pixels have the same brightness,
is assumed.

\item
Refinement of the estimate $R_* \xi$ using the information $\mathcal{U}_{\prior} = [0, 1]^{\dim \mathcal{F}}$
by the method~\eqref{eqn:reduction-estimate-in-set}
by fixed-point iteration,
i.\,e. by consecutive application of the mapping \eqref{eqn:reduction-estimate-in-set}
with $\Pi_{\mathbf{\Sigma}_{R_* \xi}} (R_* \xi)$ as the initial approximation.
We denote the obtained estimate by $\hat{u}$.

\item
Application of the sparsity-inducing transformation $T$ to $\hat{u}$.
``Sparsity-inducing'' means that
the researcher expects the chosen transform of the true transparency distribution of the object
to be sparse.

\item
\label{itm:thresholding}
Calculation of the worst-case (in $g$) variances $\sigma_{T \hat{u}}^2 = (\sigma_{(T \hat{u})_1}^2, \dots, \sigma_{(T \hat{u})_{\dim \mathcal{F}}}^2)$
of the components of $T \hat{u}$
(the diagonal matrix elements of $T \mathbf{\Sigma}_{R_* \xi} T^*$)
and calculation of $T \hat{u}_{\textnormal{thr}}$ in the following way:
$(T \hat{u}_{\textnormal{thr}})_i \bydef 0$ if $|(T \hat{u})_i| < \tau \sigma_{(T \hat{u})_i}$,
otherwise $(T \hat{u}_{\textnormal{thr}})_i \bydef (T \hat{u})_i$.

\item
Inverse transformation $T^{-1}$ of $T \hat{u}_{\textnormal{thr}}$
(if $T$ is a unitary transformation, then $T^{-1} = T^*$),
i.\,e. calculation of
$\hat{u}_{\textnormal{thr}} \bydef T^{-1} T \hat{u}_{\textnormal{thr}}$.

\item
Calculation of the projection $\Pi_{\mathbf{\Sigma}_{R_* \xi}} (\hat{u}_{\textnormal{thr}})$
that is considered to be the result of processing.
\end{enumerate}

The algorithm parameter $\tau \geq 0$
reflects a compromise between noise suppression
(the larger the value of $\tau$, the greater the noise suppression)
and distortion of images whose components are close to $0$.
As mentioned above, the step \ref{itm:thresholding}
can be considered as testing statistical hypotheses $(T U f)_i = 0$
(for the alternative $(T U f)_i \neq 0$)
for all $i$.
In this paper the criterion used in step \ref{itm:thresholding} is
based on Chebyshev's inequality:
if $(T U f)_i = 0$,
then $\Pr\left(|(T \hat{u})_i| \geq \tau \sigma_{(T \hat{u})_i}\right) \leq \tau^{-2}$ (hence, the significance level is at least $\tau^{-2}$).
Step \ref{itm:thresholding} can be also interpreted as replacement of
the original matrix $U$ with one whose kernel contains
the estimate components after the specified transform
that are affected by noise of the specified magnitude or more.

\section{Computer modeling results}
\label{sec:computer-modelling}

The results of image processing according to the described algorithm
are shown in Figs.~\ref{fig:two_slits-e0.4-l1.0}--\ref{fig:two_slits-e0.8-l2.0}.
The computer modeling was carried out for wave lengths
$(\lambda_1)^{-1} = \SI[round-mode=off, scientific-notation=false]{1.2}{\per\micro\metre}$,
$(\lambda_2)^{-1} = \SI[round-mode=off, scientific-notation=false]{0.8}{\per\micro\metre}$,
$(\lambda_3)^{-1} = \SI[round-mode=off, scientific-notation=false]{3.2}{\per\micro\metre}$,
aperture area $S_a = \SI[round-mode=off, scientific-notation=false]{25}{\square\centi\metre}$,
pixel area $S_p = \SI[scientific-notation=false]{100}{\square\micro\metre}$,
focal distance $f = \SI[scientific-notation=false]{10}{\centi\metre}$
and
the value of crystal parameter $\epsilon = \gamma/\beta$
and the dimensionless crystal length $\beta z$ indicated in figure captions,
with $\epsilon$ ranging from $0.4$ to $0.8$ and $\beta z$ ranging from $1$ to $5$.
The sensors in arms are identical ones
that are three times as large as an element of the object image.
Therefore, image processing via measurement reduction
increases resolution in addition to noise suppression.
It should be noted, however, that the objectives of superresolution
and reconstruction of the image with a small number of photons
are generally at odds with each other:
relaxing resolution requirements
allows to reconstruct the image using less photons,
as less components of the image have to be recovered.

\begin{figure}[ht]
\centering
\begin{subfigure}[t]{\subfigsize}
\centering
\includegraphics[scale=1]{{{two_slits-e0.4-l1.0-n10000000.0-src}}}
\caption{Object transparency \mbox{distribution}}
\label{fig:two-slits-e0.4-l1.0-n10000000.0-src}
\end{subfigure}
\begin{subfigure}[t]{\subfigsize}
\centering
\includegraphics[scale=1]{{{two_slits-e0.4-l1.0-n10000000.0-1}}}
\caption{The additional image at $\omega_1$}
\label{fig:two_slits-e0.4-l1.0-n10000000.0-1}
\end{subfigure}
\begin{subfigure}[t]{\subfigsize}
\centering
\includegraphics[scale=1]{{{two_slits-e0.4-l1.0-n10000000.0-2}}}
\caption{The additional image at $\omega_2$}
\label{fig:two_slits-e0.4-l1.0-n10000000.0-2}
\end{subfigure}
\begin{subfigure}[t]{\subfigsize}
\centering
\includegraphics[scale=1]{{{two_slits-e0.4-l1.0-n10000000.0-3}}}
\caption{The amplified image at $\omega_3$}
\label{fig:two_slits-e0.4-l1.0-n10000000.0-3}
\end{subfigure}
\begin{subfigure}[t]{\subfigsize}
\centering
\includegraphics[scale=1]{{{two_slits-e0.4-l1.0-n10000000.0-sum}}}
\caption{Sum of acquired images}
\label{fig:two_slits-e0.4-l1.0-n10000000.0-sum}
\end{subfigure}
\begin{subfigure}[t]{\subfigsize}
\centering
\includegraphics[scale=1]{{{two_slits-e0.4-l1.0-n10000000.0-red}}}
\caption{Reduction result, no sparsity information}
\label{fig:two-slits-e0.4-l1.0-n10000000.0-red}
\end{subfigure}
\begin{subfigure}[t]{\subfigsize}
\centering
\includegraphics[scale=1]{{{two_slits-e0.4-l1.0-n10000000.0-haar-0.3}}}
\caption{$\tau = 0.3$}
\label{fig:two_slits-e0.4-l1.0-n10000000.0-haar-0.3}
\end{subfigure}
\begin{subfigure}[t]{\subfigsize}
\centering
\includegraphics[scale=1]{{{two_slits-e0.4-l1.0-n10000000.0-haar-0.5}}}
\caption{$\tau = 0.5$}
\label{fig:two_slits-e0.4-l1.0-n10000000.0-haar-0.5}
\end{subfigure}
\begin{subfigure}[t]{\subfigsize}
\centering
\includegraphics[scale=1]{{{two_slits-e0.4-l1.0-n10000000.0-haar-0.6}}}
\caption{$\tau = 0.6$}
\label{fig:two_slits-e0.4-l1.0-n10000000.0-haar-0.6}
\end{subfigure}
\begin{subfigure}[t]{\subfigsize}
\centering
\includegraphics[scale=1]{{{two_slits-e0.4-l1.0-n10000000.0-red-s}}}
\caption{Reduction result, no sparsity information, only the image at $\omega_1$}
\label{fig:two-slits-e0.4-l1.0-n10000000.0-red-s}
\end{subfigure}
\begin{subfigure}[t]{\subfigsize}
\centering
\includegraphics[scale=1]{{{two_slits-e0.4-l1.0-n10000000.0-haar-0.6-s}}}
\caption{$\tau = 0.6$, only the image at $\omega_1$}
\label{fig:two_slits-e0.4-l1.0-n10000000.0-haar-0.6-s}
\end{subfigure}
\begin{subfigure}[t]{\subfigsize}
\centering
\includegraphics[scale=1]{{{two_slits-e0.4-l1.0-n10000000.0-haar-0.8-s}}}
\caption{$\tau = 0.8$, only the image at $\omega_1$}
\label{fig:two_slits-e0.4-l1.0-n10000000.0-haar-0.8-s}
\end{subfigure}
\caption{Processing using measurement reduction technique of parametrically amplified multiplexed images.
The scale coefficients are equalized by optical means.
Simulation was carried out for the following crystal parameters: $\epsilon = 0.4$, $\beta z = 1.0$.
The density of the photons illuminating the object $\max \langle \hat{N}_{30}(\vr)\rangle = \SI{1e7}{\per\square\centi\metre}$.
(\subref{fig:two-slits-e0.4-l1.0-n10000000.0-src})~the transparency distribution of the object,
(\subref{fig:two_slits-e0.4-l1.0-n10000000.0-1}--\subref{fig:two_slits-e0.4-l1.0-n10000000.0-3})~parametrically amplified and multiplexed acquired images and (\subref{fig:two_slits-e0.4-l1.0-n10000000.0-sum})~their sum,
(\subref{fig:two-slits-e0.4-l1.0-n10000000.0-red}--\subref{fig:two_slits-e0.4-l1.0-n10000000.0-haar-0.6})~results of their processing using the reduction technique:
(\subref{fig:two-slits-e0.4-l1.0-n10000000.0-red})~without using sparsity information and
(\subref{fig:two_slits-e0.4-l1.0-n10000000.0-haar-0.3}--\subref{fig:two_slits-e0.4-l1.0-n10000000.0-haar-0.6})~using information about sparsity in Haar transform basis;
(\subref{fig:two-slits-e0.4-l1.0-n10000000.0-red-s}--\subref{fig:two_slits-e0.4-l1.0-n10000000.0-haar-0.8-s})~the results of similar processing of only the acquired image with the best signal-to-noise ratio}
\label{fig:two_slits-e0.4-l1.0}
\end{figure}

\begin{figure}[ht]
\centering
\begin{subfigure}[t]{\subfigsize}
\centering
\includegraphics[scale=1]{{{two_slits-e0.4-l2.0-n50000.0-1}}}
\caption{The additional image at $\omega_1$}
\label{fig:two_slits-e0.4-l2.0-n50000.0-1}
\end{subfigure}
\begin{subfigure}[t]{\subfigsize}
\centering
\includegraphics[scale=1]{{{two_slits-e0.4-l2.0-n50000.0-2}}}
\caption{The additional image at $\omega_2$}
\label{fig:two_slits-e0.4-l2.0-n50000.0-2}
\end{subfigure}
\begin{subfigure}[t]{\subfigsize}
\centering
\includegraphics[scale=1]{{{two_slits-e0.4-l2.0-n50000.0-3}}}
\caption{The amplified image at $\omega_3$}
\label{fig:two_slits-e0.4-l2.0-n50000.0-3}
\end{subfigure}
\begin{subfigure}[t]{\subfigsize}
\centering
\includegraphics[scale=1]{{{two_slits-e0.4-l2.0-n50000.0-sum}}}
\caption{Sum of acquired images}
\label{fig:two_slits-e0.4-l2.0-n50000.0-sum}
\end{subfigure}
\begin{subfigure}[t]{\subfigsize}
\centering
\includegraphics[scale=1]{{{two_slits-e0.4-l2.0-n50000.0-haar-1.0}}}
\caption{$\tau = 1$}
\label{fig:two_slits-e0.4-l2.0-n50000.0-haar-1.0}
\end{subfigure}
\begin{subfigure}[t]{\subfigsize}
\centering
\includegraphics[scale=1]{{{two_slits-e0.4-l2.0-n50000.0-haar-1.5}}}
\caption{$\tau = 1.5$}
\label{fig:two_slits-e0.4-l2.0-n50000.0-haar-1.5}
\end{subfigure}
\begin{subfigure}[t]{\subfigsize}
\centering
\includegraphics[scale=1]{{{two_slits-e0.4-l2.0-n50000.0-haar-0.5-s}}}
\caption{$\tau = 0.5$, only the image at $\omega_2$}
\label{fig:two_slits-e0.4-l2.0-n50000.0-haar-0.5-s}
\end{subfigure}
\begin{subfigure}[t]{\subfigsize}
\centering
\includegraphics[scale=1]{{{two_slits-e0.4-l2.0-n50000.0-haar-0.75-s}}}
\caption{$\tau = 0.75$, only the image at $\omega_2$}
\label{fig:two_slits-e0.4-l2.0-n50000.0-haar-0.75-s}
\end{subfigure}
\caption{Processing using measurement reduction technique of parametrically amplified multiplexed images.
The scale coefficients are equalized by optical means.
Simulation was carried out for the following crystal parameters: $\epsilon = 0.4$, $\beta z = 2.0$.
The density of the photons illuminating the object $\max \langle \hat{N}_{30}(\vr)\rangle = \SI{5e4}{\per\square\centi\metre}$.
(\subref{fig:two_slits-e0.4-l2.0-n50000.0-1}--\subref{fig:two_slits-e0.4-l2.0-n50000.0-3})~parametrically amplified and multiplexed acquired images
of the object in Fig.~\ref{fig:two-slits-e0.4-l1.0-n10000000.0-src}
and (\subref{fig:two_slits-e0.4-l2.0-n50000.0-sum})~their sum,
(\subref{fig:two_slits-e0.4-l2.0-n50000.0-haar-1.0}--\subref{fig:two_slits-e0.4-l2.0-n50000.0-haar-1.5})~results of their processing
using the reduction technique and the information about sparsity in Haar transform basis;
(\subref{fig:two_slits-e0.4-l2.0-n50000.0-haar-0.5-s}, \subref{fig:two_slits-e0.4-l2.0-n50000.0-haar-0.75-s})~the results of similar processing of only the acquired image with the best signal-to-noise ratio}
\label{fig:two_slits-e0.4-l2.0}
\end{figure}

\begin{figure}[ht]
\centering
\begin{subfigure}[t]{\subfigsize}
\centering
\includegraphics[scale=1]{{{two_slits-e0.4-l5.0-n10.0-1}}}
\caption{The additional image at $\omega_1$}
\label{fig:two_slits-e0.4-l5.0-n10.0-1}
\end{subfigure}
\begin{subfigure}[t]{\subfigsize}
\centering
\includegraphics[scale=1]{{{two_slits-e0.4-l5.0-n10.0-2}}}
\caption{The additional image at $\omega_2$}
\label{fig:two_slits-e0.4-l5.0-n10.0-2}
\end{subfigure}
\begin{subfigure}[t]{\subfigsize}
\centering
\includegraphics[scale=1]{{{two_slits-e0.4-l5.0-n10.0-3}}}
\caption{The amplified image at $\omega_3$}
\label{fig:two_slits-e0.4-l5.0-n10.0-3}
\end{subfigure}
\begin{subfigure}[t]{\subfigsize}
\centering
\includegraphics[scale=1]{{{two_slits-e0.4-l5.0-n10.0-sum}}}
\caption{Sum of acquired images}
\label{fig:two_slits-e0.4-l5.0-n10.0-sum}
\end{subfigure}
\begin{subfigure}[t]{\subfigsize}
\centering
\includegraphics[scale=1]{{{two_slits-e0.4-l5.0-n10.0-haar-7.0}}}
\caption{$\tau = 7$}
\label{fig:two_slits-e0.4-l5.0-n10.0-haar-7.0}
\end{subfigure}
\begin{subfigure}[t]{\subfigsize}
\centering
\includegraphics[scale=1]{{{two_slits-e0.4-l5.0-n10.0-haar-20.0}}}
\caption{$\tau = 20$}
\label{fig:two_slits-e0.4-l5.0-n10.0-haar-20.0}
\end{subfigure}
\begin{subfigure}[t]{\subfigsize}
\centering
\includegraphics[scale=1]{{{two_slits-e0.4-l5.0-n10.0-haar-30.0}}}
\caption{$\tau = 30$}
\label{fig:two_slits-e0.4-l5.0-n10.0-haar-30.0}
\end{subfigure}
\begin{subfigure}[t]{\subfigsize}
\centering
\includegraphics[scale=1]{{{two_slits-e0.4-l5.0-n10.0-haar-7.0-s}}}
\caption{$\tau = 7$, only the image at $\omega_2$}
\label{fig:two_slits-e0.4-l5.0-n10.0-haar-7.0-s}
\end{subfigure}
\begin{subfigure}[t]{\subfigsize}
\centering
\includegraphics[scale=1]{{{two_slits-e0.4-l5.0-n10.0-haar-30.0-s}}}
\caption{$\tau = 30$, only the image at $\omega_2$}
\label{fig:two_slits-e0.4-l5.0-n10.0-haar-30.0-s}
\end{subfigure}
\caption{Processing using measurement reduction technique of parametrically amplified multiplexed images.
The scale coefficients are equalized by optical means.
Simulation was carried out for the following crystal parameters: $\epsilon = 0.4$, $\beta z = 5.0$.
The density of the photons illuminating the object $\max \langle \hat{N}_{30}(\vr)\rangle = \SI[retain-unity-mantissa=false]{3e5}{\per\square\centi\metre}$.
(\subref{fig:two_slits-e0.4-l5.0-n10.0-1}--\subref{fig:two_slits-e0.4-l5.0-n10.0-3})~parametrically amplified and multiplexed acquired images
of the object in Fig.~\ref{fig:two-slits-e0.4-l1.0-n10000000.0-src}
and (\subref{fig:two_slits-e0.4-l5.0-n10.0-sum})~their sum,
(\subref{fig:two_slits-e0.4-l5.0-n10.0-haar-7.0}--\subref{fig:two_slits-e0.4-l5.0-n10.0-haar-30.0})~results of their processing
using the reduction technique and the information about sparsity in Haar transform basis;
(\subref{fig:two_slits-e0.4-l5.0-n10.0-haar-7.0-s}, \subref{fig:two_slits-e0.4-l5.0-n10.0-haar-30.0-s})~the results of similar processing of only the acquired image with the best signal-to-noise ratio}
\label{fig:two_slits-e0.4-l5.0}
\end{figure}

\begin{figure}[ht]
\centering
\begin{subfigure}[t]{\subfigsize}
\centering
\includegraphics[scale=1]{{{two_slits-e0.8-l1.0-n10000000.0-1}}}
\caption{The additional image at $\omega_1$}
\label{fig:two_slits-e0.8-l1.0-n10000000.0-1}
\end{subfigure}
\begin{subfigure}[t]{\subfigsize}
\centering
\includegraphics[scale=1]{{{two_slits-e0.8-l1.0-n10000000.0-2}}}
\caption{The additional image at $\omega_2$}
\label{fig:two_slits-e0.8-l1.0-n10000000.0-2}
\end{subfigure}
\begin{subfigure}[t]{\subfigsize}
\centering
\includegraphics[scale=1]{{{two_slits-e0.8-l1.0-n10000000.0-3}}}
\caption{The amplified image at $\omega_3$}
\label{fig:two_slits-e0.8-l1.0-n10000000.0-3}
\end{subfigure}
\begin{subfigure}[t]{\subfigsize}
\centering
\includegraphics[scale=1]{{{two_slits-e0.8-l1.0-n10000000.0-sum}}}
\caption{Sum of acquired images}
\label{fig:two_slits-e0.8-l1.0-n10000000.0-sum}
\end{subfigure}
\begin{subfigure}[t]{\subfigsize}
\centering
\includegraphics[scale=1]{{{two_slits-e0.8-l1.0-n10000000.0-haar-0.75}}}
\caption{$\tau = 0.75$}
\label{fig:two_slits-e0.8-l1.0-n10000000.0-haar-0.75}
\end{subfigure}
\begin{subfigure}[t]{\subfigsize}
\centering
\includegraphics[scale=1]{{{two_slits-e0.8-l1.0-n10000000.0-haar-2.0}}}
\caption{$\tau = 2$}
\label{fig:two_slits-e0.8-l1.0-n10000000.0-haar-2.0}
\end{subfigure}
\begin{subfigure}[t]{\subfigsize}
\centering
\includegraphics[scale=1]{{{two_slits-e0.8-l1.0-n10000000.0-haar-1.0-s}}}
\caption{$\tau = 1$, only the image at $\omega_1$}
\label{fig:two_slits-e0.8-l1.0-n10000000.0-haar-1.0-s}
\end{subfigure}
\begin{subfigure}[t]{\subfigsize}
\centering
\includegraphics[scale=1]{{{two_slits-e0.8-l1.0-n10000000.0-haar-2.0-s}}}
\caption{$\tau = 2$, only the image at $\omega_1$}
\label{fig:two_slits-e0.8-l1.0-n10000000.0-haar-2.0-s}
\end{subfigure}
\caption{Processing using measurement reduction technique of parametrically amplified multiplexed images.
The scale coefficients are equalized by optical means.
Simulation was carried out for the following crystal parameters: $\epsilon = 0.8$, $\beta z = 1.0$.
The density of the photons illuminating the object $\max \langle \hat{N}_{30}(\vr)\rangle = \SI{1e7}{\per\square\centi\metre}$.
(\subref{fig:two_slits-e0.8-l1.0-n10000000.0-1}--\subref{fig:two_slits-e0.8-l1.0-n10000000.0-3})~parametrically amplified and multiplexed acquired images
of the object in Fig.~\ref{fig:two-slits-e0.4-l1.0-n10000000.0-src}
and (\subref{fig:two_slits-e0.8-l1.0-n10000000.0-sum})~their sum,
(\subref{fig:two_slits-e0.8-l1.0-n10000000.0-haar-0.75}--\subref{fig:two_slits-e0.8-l1.0-n10000000.0-haar-2.0})~results of their processing
using the reduction technique and the information about sparsity in Haar transform basis;
(\subref{fig:two_slits-e0.8-l1.0-n10000000.0-haar-1.0-s}, \subref{fig:two_slits-e0.8-l1.0-n10000000.0-haar-2.0-s})~the results of similar processing of only the acquired image with the best signal-to-noise ratio}
\label{fig:two_slits-e0.8-l1.0}
\end{figure}

\begin{figure}[ht]
\centering
\begin{subfigure}[t]{\subfigsize}
\centering
\includegraphics[scale=1]{{{two_slits-e0.8-l2.0-n100000.0-1}}}
\caption{The additional image at $\omega_1$}
\label{fig:two_slits-e0.8-l2.0-n100000.0-1}
\end{subfigure}
\begin{subfigure}[t]{\subfigsize}
\centering
\includegraphics[scale=1]{{{two_slits-e0.8-l2.0-n100000.0-2}}}
\caption{The additional image at $\omega_2$}
\label{fig:two_slits-e0.8-l2.0-n100000.0-2}
\end{subfigure}
\begin{subfigure}[t]{\subfigsize}
\centering
\includegraphics[scale=1]{{{two_slits-e0.8-l2.0-n100000.0-3}}}
\caption{The amplified image at $\omega_3$}
\label{fig:two_slits-e0.8-l2.0-n100000.0-3}
\end{subfigure}
\begin{subfigure}[t]{\subfigsize}
\centering
\includegraphics[scale=1]{{{two_slits-e0.8-l2.0-n100000.0-sum}}}
\caption{Sum of acquired images}
\label{fig:two_slits-e0.8-l2.0-n100000.0-sum}
\end{subfigure}
\begin{subfigure}[t]{\subfigsize}
\centering
\includegraphics[scale=1]{{{two_slits-e0.8-l2.0-n100000.0-haar-1.0}}}
\caption{$\tau = 1$}
\label{fig:two_slits-e0.8-l2.0-n100000.0-haar-1.0}
\end{subfigure}
\begin{subfigure}[t]{\subfigsize}
\centering
\includegraphics[scale=1]{{{two_slits-e0.8-l2.0-n100000.0-haar-1.5}}}
\caption{$\tau = 1.5$}
\label{fig:two_slits-e0.8-l2.0-n100000.0-haar-1.5}
\end{subfigure}
\begin{subfigure}[t]{\subfigsize}
\centering
\includegraphics[scale=1]{{{two_slits-e0.8-l2.0-n100000.0-haar-1.0-s}}}
\caption{$\tau = 1$, only the image at $\omega_2$}
\label{fig:two_slits-e0.8-l2.0-n100000.0-haar-1.0-s}
\end{subfigure}
\begin{subfigure}[t]{\subfigsize}
\centering
\includegraphics[scale=1]{{{two_slits-e0.8-l2.0-n100000.0-haar-1.5-s}}}
\caption{$\tau = 1.5$, only the image at $\omega_2$}
\label{fig:two_slits-e0.8-l2.0-n100000.0-haar-1.5-s}
\end{subfigure}
\caption{Processing using measurement reduction technique of parametrically amplified multiplexed images.
The scale coefficients are equalized by optical means.
Simulation was carried out for the following crystal parameters: $\epsilon = 0.8$, $\beta z = 2.0$.
The density of the photons illuminating the object $\max \langle \hat{N}_{30}(\vr)\rangle = \SI{1e5}{\per\square\metre}$.
(\subref{fig:two_slits-e0.8-l2.0-n100000.0-1}--\subref{fig:two_slits-e0.8-l2.0-n100000.0-3})~parametrically amplified and multiplexed acquired images
of the object in Fig.~\ref{fig:two-slits-e0.4-l1.0-n10000000.0-src}
and (\subref{fig:two_slits-e0.8-l2.0-n100000.0-sum})~their sum,
(\subref{fig:two_slits-e0.8-l2.0-n100000.0-haar-1.0}, \subref{fig:two_slits-e0.8-l2.0-n100000.0-haar-1.5})~results of their processing
using the reduction technique and the information about sparsity in Haar transform basis;
(\subref{fig:two_slits-e0.8-l2.0-n100000.0-haar-1.0-s}, \subref{fig:two_slits-e0.8-l2.0-n100000.0-haar-1.5-s})~the results of similar processing of only the acquired image with the best signal-to-noise ratio}
\label{fig:two_slits-e0.8-l2.0}
\end{figure}

One can see that additional information about sparsity
enables higher noise suppression
without compromising reducing obtained resolution too much.
Increase of $\tau$ leads to better noise suppression
(cf., e.\,g., Figs.~\ref{fig:two_slits-e0.4-l1.0-n10000000.0-haar-0.3} and \ref{fig:two_slits-e0.4-l1.0-n10000000.0-haar-0.5}),
but also worse distortions
caused by discarding ``significant'' image components as well
(cf., e.\,g., Figs.~\ref{fig:two_slits-e0.4-l2.0-n50000.0-haar-1.0} and \ref{fig:two_slits-e0.4-l2.0-n50000.0-haar-1.5}).
Too large values of $\tau$
cause degradation of image fidelity
due to distortion outweighing improved noise suppression,
as small-scale image details are suppressed as well.
Therefore,
one should choose the maximal value of $\tau$
that preserves the details of interest.
To do that, one can model processing of a test image
that contains the required details
and choose the largest value of $\tau$
that preserves them,
or visually compare the reduction results for different $\tau$
and select its final value by ternary search.

The transform whose result for the transparency distribution of the object
is sparse
that is usually employed in image processing by the means of compressed sensing
is discrete cosine transform (DCT)
\cite{chan_lu_2014, imaging_small_n_photons, mertens_et_al_2017}.
In \cite{sparsity_property_influence},
several transforms (identity transform, discrete wavelet transform and DCT)
were reviewed and the advantages of DCT were shown.
However, in \cite{gi_sparsity_eng}
it was shown that Haar transform may be preferable
in the case of a transparency distribution
that contains areas of weakly changing transparency with sharp borders
if these areas are large compared to the resolution of the ideal MT
and the location of the borders is important to the researcher,
since those features align well with the vectors of the Haar transform basis.
As the object images in Figs.~\ref{fig:two_slits-e0.4-l1.0}--\ref{fig:two_slits-e0.8-l2.0} are of this type,
the sparsity-inducing transform used in these figures is the Haar transform.

In all these figures
results of processing of \emph{all} images
are compared to results of processing of a \emph{single} image,
namely, the one with the best signal-to-noise ratio.
Since processing only a single image means that
some of the obtained measurements are discarded,
this results in worse estimate quality,
even if the sum of acquired images is not noticeably different from the best image.
For example, in Fig.~\ref{fig:two_slits-e0.4-l5.0}
one can see that the same values of $\tau$ cause more distortion
when processing a single image.
In Fig.~\ref{fig:two_slits-e0.4-l5.0-n10.0-haar-30.0-s}
only a single component, corresponging to uniform transparency,
remains in the result of processing a single image,
while the same value of $\tau$ produces an acceptable,
although suboptimal, image shown in Fig.~\ref{fig:two_slits-e0.4-l5.0-n10.0-haar-30.0}.
Furthermore, in the general case of additional noise
multiplexing provides the means for further noise suppression
if noise photons in different arms are detected independently.

As far as the parameters of the ANPC are concerned, larger crystal length $\beta z$
leads to more amplification and more photons.
The improvement of the number of photons is especially large when $\beta z$
is increased from $1$ to $2$.
The results do not depend on the value of the parameter $\epsilon$ as much.

\section*{Conclusion}

The problem of recovering images acquired in photon-sparse conditions
can be solved by taking advantage of the additional information available to the researcher
about the measurement process and about the object.
Alternatively one can make the detection conditions worse (e.\,g. use sensors with less resolution)
while preserving the same estimation quality.
In this work, the additional information about the object is
the information that the object transparency distribution is not arbitrary,
namely, transparencies of close pixels tend to be close.
This information is formalized as sparsity
of the result of a given transform of the transparency distribution,
similar to compressed sensing.

In compressed sensing, as a rule, the measurement error
is modeled as an arbitrary vector with bounded norm.
In the proposed method it is modeled as a random vector,
and selection of the estimate components which are considered to be zero
is based on the statistical properties of the estimate components, namely, their variances.
The use of covariances of the estimate components in addition to their variances
is a subject of further research.
Another subject of further research are the opportunities provided by multiplexing
for analysis of the measurement data,
e.\,g., for verifying the reliability
of both the measurement model \cite{pytyev_chulichkov_1991, pytyev_serdobolskaya_1988}
and the results of reduction \cite{pytyev_1993}.

We consider that computer modeling based on the developed algorithm
showed high efficiency of the developed reduction technique
for parametric amplification of images and frequency conversion
in the sense of improvement of both their quality and their noise immunity.

Finally, we emphasize once again that
we are dealing with parametric interaction with low-frequency pumping.
The carrier wavelength of the original image may be in the ultraviolet range,
while the pump frequency may belong to the visible range.
Further applications of the measurement reduction technique
to processing of quantum images
are being developed.
This work was supported by Russian Foundation for Basic Research, grant 18-01-000598 A.

\section*{Acknowledgments}

The authors acknowledge discussions with Prof. A.\,V.\,Belinsky. 

\bibliographystyle{unsrt}
\bibliography{pc_bibl}
\end{document}